\documentclass[prb,twocolumn,superscriptaddress,floatfix,showpacs]{revtex4}
\usepackage{amsmath,amsfonts,epsf,dcolumn,natbib}
\usepackage{ascii}
\newcommand{\half}{\tfrac{1}{2}}

\newcommand{\be}{\begin{equation}}
\newcommand{\ee}{\end{equation}}
\newcommand{\bea}{\begin{eqnarray}}
\newcommand{\eea}{\end{eqnarray}}
\begin{document}
\bibliographystyle{plainnat}

\title{Issues and Challenges in Orbital-free Density Functional Calculations}

\author{V.V.~Karasiev}
\email{vkarasev@qtp.ufl.edu}
\author{S.B.~Trickey}
\email{trickey@qtp.ufl.edu}
\affiliation{Quantum Theory Project, 
Departments of Physics and of Chemistry, P.O. Box 118435, 
University of Florida, Gainesville FL 32611-8435}


%
\begin{abstract}
Solving the Euler equation which corresponds to the
energy minimum of a density functional expressed in orbital-free
form involves related but distinct computational challenges. 
One is the choice between all-electron and pseudo-potential 
calculations and, if the latter, construction of the pseudo-potential.
Another is the stability, speed, and accuracy of solution algorithms.   
Underlying both is the fundamental issue of satisfactory quality of
the approximate functionals (kinetic energy and exchange-correlation).
We address both computational issues and illustrate them by some
comparative performance testing of our recently developed modified-conjoint
generalized gradient approximation kinetic energy functionals. Comparisons
are given for atoms, diatomic molecules, and some simple solids.
\end{abstract}


\maketitle

\renewcommand{\baselinestretch}{1.05}\rm

\section{Introduction}

Investigation of orbital-free density functional theory (OF-DFT) 
\cite{WangCarter2000,%
LudenaKarasiev2002,Zhou06,Garcia0807,Perdew07,GarciaCervera08, %
Ghiringhelli08,Eek06,Perspectives,Karasiev..Harris.2009,%
Trickey..Jones.2009,HuangCarter10},
including development of approximate orbital-free kinetic energy 
(OFKE) functionals, has at least two motivations.  One is simply the 
beguiling notion of direct realization of the content of the Hohenberg-Kohn 
theorem 
\cite{Hohenberg-Kohn,Yang-Parr-book,DreizlerGrossBook,KryachkoLudena,Eschrig}.
The other is practical, namely the possibility of eliminating the 
computational bottleneck  
of solving the Kohn-Sham (KS) eigenvalue equations, thereby dramatically
broadening the applicability of Born-Oppenheimer
molecular dynamics run with DFT electronic energies.  The practical
aspect is the main focus of the present work.  

In OF-DFT, the total electronic energy of an $N_e$ electron system  
is a functional of the electron density $n({\mathbf r})$
\begin{eqnarray}
E^{\rm OF\mbox{-}DFT}[n]&=&T_{\rm s}[n]+E_{\rm Ne}[n]+E_{\rm H}[n] \nonumber \\ [8pt]
&&+\,E_{\rm xc}[n] + E_{\rm NN},
\label{A2}
\end{eqnarray}
where $T_{\rm s}[n]$ is the KS (non-interacting) kinetic energy functional
given explicitly as a density functional,  
$E_{\rm Ne}[n]$ is the nuclear-electron interaction energy, 
$E_{\rm H}[n]$ is the Hartree energy (classical
electron-electron repulsion), $E_{\rm xc}[n]$ is the exchange-correlation (XC)
energy functional, and $E_{\rm NN}$ is the inter-nuclear 
repulsion energy. Minimization of the functional
Eq.\ (\ref{A2}) gives a single Euler equation to be solved,
\begin{equation}
\frac{\delta T_{\rm s}[n]}{\delta n({\bf r})}+v_{\rm KS}([n];{\bf r})=\mu\, .
\label{A3}
\end{equation}
Here $v_{\rm KS}$ is the Kohn-Sham potential, 
 $\delta (E_{\rm Ne} + E_{\rm H} + E_{\rm xc})/\delta n$ and $\mu$ is the chemical
potential.  The ordinary KS
equation has the same potential but requires solution for $N_e$ or
$N_e/2$ orbitals (in the all-electron, spin-polarized and
non-spin-polarized cases respectively).  Solution of the
ordinary KS problem scales computationally as $\approx N_e^3$ 
in general, whereas
solution of Eq.\ (\ref{A3}) should scale approximately linearly.  

Practical implementation of OF-DFT requires approximation
of both $T_{\rm s}[n]$ and $E_{\rm xc}[n]$.  Simply because of their
relative magnitudes, the quality of an OF-DFT calculation
is dominated by the quality of the approximate $T_{\rm s}$.  There
are two distinct classes of approximation in the literature,
one-point functionals,
\be
T_{\mathrm s}[n] = \int  t_{\mathrm s}([ n];{\mathbf r}) d^3{\mathbf r}
\label{onepoint}
\ee
and two-point functionals
\be
T_{\mathrm s}[n] = \int  f_{1,\mathrm s} ([ n];{\mathbf r})
\chi({\mathbf r},{\mathbf r}^\prime) %
f_{2,\mathrm s}([ n];{\mathbf r}^\prime) d^3{\mathbf r}d^3{\mathbf r}^\prime \; .
\label{twopoint}
\ee
Here $f_{1,\mathrm s}$ and $f_{2,\mathrm s}$ are weighting functionals
and $\chi({\mathbf r},{\mathbf r}^\prime) $ is a type of response
function.  For reasons of computational efficiency as well as conceptual
simplicity (two-point functionals take the development out of
the framework of an effective Kohn-Sham equation (see Eq.\ (\ref{II3}) below) 
unless an
optimized effective potential\cite{OEP} is used, itself an extra 
complication), we (and our collaborators) have focused exclusively 
on one-point functionals and do so here as well. 

An interesting feature of the literature on developing approximate
OFKE functionals, including our contributions with collaborators, is
that there are more tests of approximate functionals using inputs from
other sources ({\it e.g.} conventional KS calculations, Hartree-Fock
calculations, etc.) than tests by solving the Euler equation,
Eq.\ (\ref{A3}).  A side effect is that comparatively little is known
about the difficulty of solving that equation with approximations
other than of the Thomas-Fermi kind (see below) and about the relative
effectiveness of various solution techniques.

To frame that issue and the calculations reported here, it is 
useful to decompose the non-interacting KE functional into the von Weizs\"acker
contribution\cite{Weizsacker} plus a non-negative remainder, the Pauli term 
\cite{TalBader78,BartolottiAcharya82,%
Harriman87,LevyOu-Yang88}, 
\begin{equation}
T_{\rm s}[n]=T_{\rm W}[n]+T_{\theta}[n], ~T_{\theta}[n] \; \geq 0  \; .   
\label{B7}
\end{equation}
The von Weizs\"acker functional (in Hartree atomic units) is 
\begin{equation}
T_{\rm W}[n] = 
\frac{1}{8}\int  \frac{|\nabla n({\bf r})|^2}
 {n({\bf r})}  d^3{\bf r} \equiv \int  t_{\rm W} ([ n];{\mathbf r})  d^3{\bf r}
\,.
\label{B1a}
\end{equation}
It is exact for one electron and for a two-electron singlet.  From
\be
\frac{\delta T_{\rm W}[n]}{\delta n({\bf r})}=
\frac{1}{\sqrt{n({\bf r})}}(-\frac{1}{2}\nabla^2)\sqrt{n({\bf r})}   \; ,
\label{II5}
\ee
the Euler equation Eq.\ (\ref{A2}) takes a Schr\"odinger-like form
\cite{DebGhosh83,Levy..Sahni84,LevyOu-Yang88}
\be
\left\{-\frac{1}{2} \nabla^2 + v_{\theta}([n];{\bf r})+
v_{\rm KS}([n];{\bf r})\right\}\sqrt{n({\bf r})}=\mu \sqrt{n({\bf r})} \; .
\label{II3}
\ee
Observe that, unlike familiar quantum mechanical eigenvalue 
problems, the ``orbital'' in Eq.\ (\ref{II3}) is normalized to 
$N_e$, not unity.
Here $v_{\theta}$ is the Pauli potential,
\begin{align}
v_{\theta}([n];{\bf r})= &\frac{\delta T_{\theta}[n]}{\delta n({\bf r})} %
\nonumber \\
v_{\theta}([n];{\bf r})\ge & 0\, .
\label{II4}
\end{align}
Non-negativity of $T_\theta$ and $v_\theta$ has proved to be an 
important pair of constraints for OFKE functional development 
\cite{Perspectives,Karasiev..Harris.2009,Trickey..Jones.2009}.

Eq.\ (\ref{II3}) resembles the ordinary
KS equation, a fact that has led to contradictory statements about solution
techniques.  On the one hand, Ref.\ \onlinecite{Levy..Sahni84}
declares that Eq.\ (\ref{II3}) ``\ldots can be solved iteratively to
self-consistency by any Kohn-Sham computer program: just select the
lowest eigenvalue.  The solution is very simple and quick, for there
is only {\it one} `orbital' \ldots ''.  Ref.\ \onlinecite{Chan..Handy.2001}  
makes precisely the contrary claim, at least in the context of the widely 
used Gaussian-type orbital (GTO) basis sets.  Those authors 
expanded $\sqrt n$ in a 
GTO basis with coefficients $c_i$, with respect to which they minimized  
${\mathcal L}:= E^{\rm OF\mbox{-}DFT}[n] - \mu N_e$.  
They state that ``Due to the highly nonquadratic nature of 
the kinetic energy, the optimization of $\mathcal L$ is a nontrivial problem.  
The iterative self-consistent procedure used in Kohn-Sham calculations does 
not work, and we require more robust minimization techniques. Moreover, 
\ldots first derivative methods such as conjugate gradient minimization and 
quasi-Newton search perform poorly, requiring many hundreds of iterations to 
achieve convergence.'' A related discussion and references to the few
earlier papers on the issue is at p.\ 135 of Ref.\ \onlinecite{Yang-Parr-book}.
This is one of the issues addressed in the present study. 

\section{Approximate Kinetic Energy Functionals}

To set the stage for another technical issue, we consider types of
approximate one-point OFKE functionals next.  
For work on minimization involving 
two-point functionals, see Refs.\ \onlinecite{Ho..Carter08}, 
\onlinecite{Hung..Carter10} and references therein.

\subsection{Thomas-Fermi Type}

Diverse approximate OFKE functionals can be written in the generic form
\begin{align}
T_{\rm s}[n] =& T_{\rm W}[n] + \lambda T_{\rm TF} [n] + T_\Delta[n] \nonumber \\
& 0 \le \lambda \le 1   \; .
\label{genericOFKE}
\end{align}
The simplest local approximation for the KE is the Thomas-Fermi (TF) 
\cite{Thomas,Fermi} functional 
\begin{align}
T_{\rm TF}[n]\equiv & \int t_{\rm TF}([n];{\bf r})\,d^3{\bf r}
= c_0\int n^{5/3} ({\bf r})\, d^3{\bf r} \nonumber \\
c_0 =& \tfrac{3}{10}(3\pi^2)^{2/3} 
\label{I6}
\end{align}
alone.  The approximation  $T_\Delta=0$ 
and $\lambda = 1$ is widely used in many OF-DFT applications
(see Ref.\ \onlinecite{LudenaKarasiev2002} for discussion and references) 
despite its known deficiencies \cite{Lieb81}. 
A related form, commonly called 
Thomas-Fermi-Dirac-von Weizs\"acker theory, is a linear 
combination of $T_{\rm TF}$ with some fraction of $T_{\rm W}$,
\be
T_{\rm TFvW,\alpha}= \alpha T_{\rm W} + T_{\rm TF} \; ; \; 0 \le \alpha \le 1
\label{TTFvW}
\ee
along with the local Dirac exchange functional. 
Early reports of special self-consistent OF-DFT calculations 
mentioned above were for this model 
\cite{Tomishima.Yonei.1966,Yang.1986,Chan..Handy.2001}.  

As an aside, there is an extensive literature of efforts to determine
an optimal value of $\alpha$ in Eq.\ (\ref{TTFvW}).  Since the Pauli
term decomposition, Eq.\ (\ref{B7}), provides both an exact lower
bound on $T_{\rm s}$ and leads to the density Schr\"odinger equation,
Eq.\ (\ref{II3}), that decomposition, and its elaboration
Eq.\ (\ref{genericOFKE}), seems preferable to using $T_{\rm TFvW,\alpha}$
and attempting to optimize $\alpha$.  But because $T_{\rm TFvW,\alpha}$ is
prevalent in the literature, we  consider the numerical
issues associated with it as well.
   
\subsection{Generalized Gradient Approximation KE Functionals}

Generalized gradient approximations (GGA) are best known in DFT as 
improvements on the local approximation for $E_{\rm xc}$.  
For either $E_{\rm xc}$ or 
$T_{\rm s}$, a GGA is a
truncation of the corresponding gradient expansion which is 
altered to meet relevant constraints and suppress
unphysical behaviors.  For the KE functional, a GGA can be written as
\be
T_{\rm s}^{\rm GGA}[n]=\int   t_{\rm TF}([n];{\bf r}) F_{\rm t}(s({\bf r}))d^3{\mathbf r} \; ,
\label{III1}
\ee
where $F_{\rm t}$ is the kinetic energy enhancement factor. It is 
a function of the dimensionless reduced density gradient, 
\be
s \equiv  \frac{|\nabla n|}{(2k_F)n}
 =  \frac{1}{2(3\pi^2)^{1/3}} \, \frac{|\nabla n|}{n^{4/3}}\,. 
\label{sdefn}
\ee
Because $t_{\rm W}=\frac{5}{3}s^2 t_{\rm TF}$, the GGA Pauli term in 
Eq.\ (\ref{B7}) is 
\begin{align}
T_{\theta}^{\rm GGA}[n]= & \int  t_{\rm TF}([n];{\bf r}) %
F_{\theta}(s({\bf r}))d^3{\mathbf r}  \nonumber \\
F_{\theta}(s)= & F_{\rm t}(s)-\frac{5}{3} s^2 \; .
\label{III2}
\end{align}

\begin{figure}        
\epsfxsize=7.4cm
\epsffile{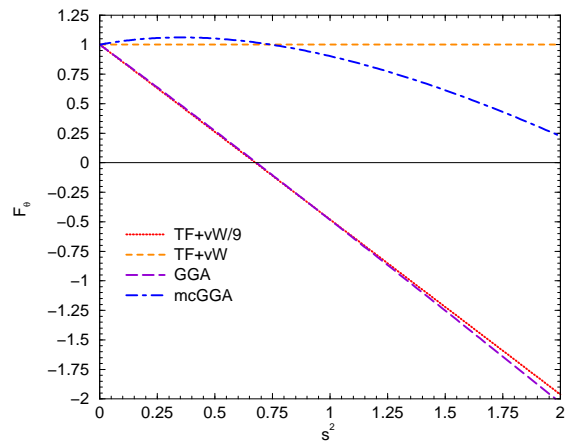}
\caption{
Pauli term enhancement factors $F_{\theta}$ of OFKE functionals 
as a function of $s^2$. GGA denotes the Tran-Wesolowski functional,
mcGGA denotes the PBE2 functional.  See text for details. 
}
\label{fig1}
\end{figure}

Ref.\ \onlinecite{Karasiev..Harris.2009} showed that the KS KE of 
a molecular system is dominated by the behavior
of $F_{\theta}$ over a relatively small range of $s$.  For much
of that range, Figure
\ref{fig1} displays the Pauli enhancement factors for the functionals 
$T_{\rm TFvW,\alpha}$, with $\alpha = 1, 1/9$  Eq.\ (\ref{TTFvW}),   
the Tran-Wesolowski GGA \cite{TranWesolowski02} (PBE-TW), and 
the mcGGA functional (PBE2) of Ref.\ \onlinecite{Perspectives}.  
The latter two use the same 
enhancement factor form as the Perdew, Burke, and Ernzerhof
(PBE) \cite{PBE} GGA X functional,  
$F_{\rm x}(s)=1+cs^2/(1+as^2)$.  
In PBE-TW  $F_{\rm t} \propto F_{\rm x,PBE}$ with parameters fitted 
to reproduce the kinetic energy of a small training set, 
an assumption  
called conjointness. PBE2 is a ``modified conjoint'' GGA (mcGGA) 
functional because 
the parameters in it were constrained to satisfy Pauli-term non-negativity;
see Ref.\ \onlinecite{Karasiev..Harris.2009} for details.

Observe in Fig.\ \ref{fig1} that the PBE-TW and $T_{\rm TFvW,\alpha=1/9}$ Pauli
enhancement factors are almost identical, especially for small 
 $s$.  There, both have negative slope  (with
respect to $s^2$) which causes violation of $v_\theta$ 
non-negativity \cite{Karasiev..Harris.2009}, recall Eq.\ (\ref{II4}).  
The common property of the 
$T_{\rm TFvW,\alpha=1}$  and PBE2 approximations is satisfaction of 
that non-negativity constraint.  The low slope of the
PBE2 enhancement factor at small values of $s^2$, 
$F_{\theta}^{\rm PBE2}(s)\approx 1+0.3642s^2$, makes the enhancement 
factors for $T_{\rm TFvW,\alpha=1}$
and $T_{\rm PBE2}$ close for $s<1$.  This comparison suggests that the
results obtained with the PBE-TW KE functional should be close to
those from $T_{\rm TFvW,\alpha=1/9}$ and, similarly, the results 
from PBE2 should be close to those from $T_{\rm TFvW,\alpha=1}$.

A technical problem common to these GGAs is that both 
$v_{\theta,\rm PBE-TW}$ and $v_{\theta,\rm PBE2}$ are singular at nuclear sites,
the former negative, the latter positive.  Numerical solution
of the Euler equation Eq.\ (\ref{A3}) must address this problem, 
an issue to which we return below.  

\section{All-electron Solutions of the OF-DFT Euler Equation}

As in ordinary KS calculations, solution of
Eq.\ (\ref{II3}) can be either all-electron or via pseudo-potentials.
In this section, we consider all-electron solutions, 
by both GTO-basis and numerical grid techniques and address
pseudo-potentials in the subsequent section.  

\subsection{Atoms}

To test the notion that any standard KS code can be used straightforwardly 
\cite{Levy..Sahni84}, we modified the GTO-basis code SOAtom to
handle $T_\theta$ and $v_\theta$ as in Eqs.\ (\ref{B7}), (\ref{II3}),
and (\ref{II4}).  SOAtom, a part of the GTOFF suite
\cite{Boettger98a,TAB}, solves the KS equation in a Hermite Gaussian
basis with analytical evaluation of all the matrix elements except for
those involving XC.  Those are done on a radial grid.  We also
modified the Laaksonen all-numerical diatomic molecular code
\cite{Laak} correspondingly.  It is based on a prolate spheroidal
grid.  

Insofar as numerical stability is concerned, the results are quite
clear.  Even for $T_{\rm TFvW,\alpha=1}$ with simple Slater exchange ({\it
  i.e.} TFvWD), the typical iterative SCF procedure is only marginally
stable.  The problem is worse in the GTO basis than in the grid-based
calculation, at least in the specific sense that a simple SCF
stabilization procedure (Pratt, {\it i.e.} linear mixing of a 
fraction of current iteration density and the rest from the 
preceding iteration) fails completely for many OF-DFT calculations. 
Ordinary KS calculations on the same atoms with the same simple 
stabilization scheme converge in a few iterations.  

The lithium and carbon atoms are examples.
For Li in a 9s GTO basis in the SOAtom code, the pure TF
form ($T_{\rm s}=T_{\rm TF}$, {\it i.e.}, Eq.\ (\ref{TTFvW}) with $\alpha =0$),
the $T_\theta = T_{\rm TF}$ form (Eq.\ (\ref{genericOFKE}) with $\lambda
=1$, $T_\Delta =0$), and the $T_\theta = \lambda T_{\rm TF}$ form
(Eq.\ (\ref{genericOFKE}) with $T_\Delta =0$) can be brought to numerical convergence but
the mcGGA form $T_\theta = T_{\rm mcGGA}-T_{\rm vW}$ cannot. For the successes, more
iterations by one to two orders of magnitude are required than for
conventional KS and the numerical convergence is poor.  One can get to
fractional total energy errors of $10^{-4} \rightarrow 10^{-2}$ for
lighter to heavier atoms respectively, but not much better.  The
contrast with conventional KS atomic calculations is stark: in them
convergence to $10^{-6}$ is trivial to achieve.  

Table \ref{tab:TABLE1} 
illustrates this point with comparison of numerical grid and 
GTO-basis results 
for  $T_{\rm s}=T_{\rm vW}$ and $T_{\rm vW}+T_{\rm TF}$ with
simple Slater exchange (Dirac
exchange) on H, Li, and Ne. (The GTO calculations are with 9s basis sets
for H and Li, 13s for Ne.)  The two total energies for Li differ at
the 1 mHartree scale. Notice that the numerical-grid results match
rather well with the values from Ref.\ \onlinecite{Chan..Handy.2001},
which were calculated with a direct minimization scheme, not a modified
KS code.  Ironically, a misbehavior of simple Slater
exchange, namely that it satisfies the virial theorem in the form
$E_{\rm tot}= - T_{\rm s}$ (which the exact $E_{\rm xc}$ does not), in this case
highlights the convergence problem, especially in the GTO calculation.

Results for the carbon atom in the GTO basis, not shown in the Table, 
are worse.  The $T_{\rm TFvW,\alpha=1/5}$ calculation with a 13s basis 
cannot be brought to SCF convergence, even with tricks such as starting 
with full
$T_{\rm W}$ and no $T_{\rm TF}$ contribution, then slowly scaling down the
former while scaling up the latter.  The corresponding standard KS
calculation converges trivially.


\begin{table*}[tb]
\caption{\label{tab:TABLE1}
All-numerical and GTO results for the atoms H, Li, and Ne for
the $T_{\rm TFvW,\alpha=0,1}$ models with simple Slater exchange.
Energies in Hartree a.u.
} 
\begin{ruledtabular}
\begin{tabular}{lrrrrr}
  & $T_{\rm s}= T_{\rm W}$ Numer. & $T_{\rm s}= T_{\rm W}$ GTO & $T_{\rm s}= T_{\rm W} + T_{\rm TF}$ Numer.  &$T_{\rm s}= T_{\rm W} + T_{\rm TF}$ GTO & $T_{\rm s} = T_{\rm W} + T_{\rm TF}$ \footnotemark \\
\hline
\hspace*{10pt}\\ [-8pt]
H Atom &  &   &  & &\\
$E_{\rm tot}$ & -0.406534 & -0.400737  &-0.261827   & -0.259969 & -0.2618 \\
$T_{\rm s}$   &  0.406534 & 0.859699   & 0.261827  &0.262042 & --- \\
$T_{\theta}$ &0.000 & 0.000   & 0.091034  & 0.090221 & --- \\
$\mu$ & -0.1943 &-0.1764   & -0.0715   &-0.0696 & -0.071 \\
\hline
Li Atom &  &   &  & &\\
$E_{\rm tot}$ &-8.525825 & -8.523413  &-4.105425  & -4.096347 & -4.1054 \\
$T_{\rm s}$ &8.525825 & 8.523126   & 4.105425  & 4.103660 & --- \\
$T_{\theta}$ &0.000 & 0.000   & 2.019249  & 2.009622 & ---  \\
$\mu$ &-0.9575 & -0.9526 &-0.1306  &-0.0.1365 & -0.131 \\
\hline
Ne Atom &  &   &  & &\\
$E_{\rm tot}$ & -274.68080 & -274.652253  &-85.734451  & -85.730041 &-85.7343 \\
$T_{\rm s}$   &  274.68080 & 274.664688  & 85.734438  & 85.728273 & --- \\
$T_{\theta}$ &0.000 & 0.000   & 54.352106  & 54.347495 & --- \\
$\mu$ &-7.0607 & -7.0594 &-0.1807  &-0.1806 & -0.181\\
\end{tabular}
\end{ruledtabular}
\footnotetext[1] {From Ref.\ \onlinecite{Chan..Handy.2001}}
\end{table*}


For the numerical-grid calculations, SCF convergence is very slow
compared to standard KS calculations, but reasonable results can be
obtained.  Table \ref{tab:table2} shows total energies for the first
row atoms obtained from numerical-grid self-consistent OF-DFT
calculations with various OFKE approximations, again with Slater
exchange. For TFvWD and
$T_{\rm TFvW,\alpha=1/5,1/9}$, comparison with the direct minimization of
Ref.\ \onlinecite{Chan..Handy.2001} (the first six columns of data) 
confirms that our calculations succeeded.  

Note that the total energies from the TF+vW and mcGGA(PBE2) kinetic
energy functionals are overestimated (as a consequence of
overestimation of the KS KE).  In contrast, all of the 
functionals with scaled von Weizs\"acker contributions
{\it underestimate} the KS KE, so that the resulting total energies
are {\it below}  the
reference KS values.  Such behavior is characteristic of 
a failure of $N$-representability in the KE functional \cite{Ayers07}. 
Observe also that TF+1/9vW and
GGA(PBE-TW) total energies are close to each other, though the 
functional forms differ. 

Table \ref{tab:table3} shows the effects of using the full
LDA $E_{\rm xc}$, in this case the VWN parameterization \cite{VWN}.
Unsurprisingly but reassuringly, inclusion of the {\it C} functional 
shifts the total energies downward without altering the trends.

\begin{table*}[tb]
\caption{\label{tab:table2}
Self-consistent atomic total energies obtained 
from various OFKE functionals (Hartree a.u.) and simple Slater exchange.
} 
\begin{ruledtabular}
\begin{tabular}{lccccccccc}
 & 1/9 vW+TF\footnotemark[1] & 1/9 vW+TF & 1/5 vW+TF\footnotemark[1] & 1/5 vW+TF &  vW+TF\footnotemark[1] &  vW+TF & GGA (PBE-TW) & mcGGA (PBE2) & KS\footnotemark[2]~ \\
\colrule
H  & -0.6664    & -0.6664 & -0.5666 & -0.5666 & -0.2618  & -0.2618  & -0.71 & -0.32 &-0.4065 \\
He & -3.2228    & -3.2228 & -2.8184 & -2.8184 & -1.4775  & -1.4775  &-3.4 &-1.5&-2.7236\\
Li & -8.2515    & -8.2515 & -7.3227 & -7.3227 & -4.1054  & -4.1054  &-8.6 &-4.1 &-7.1749\\
Be & -16.1631   & -16.1631 & -14.4841 & -14.4841 & -8.4922  & -8.4922 &-16.7 &-8.4&-14.2233\\
B  & -27.2876   & -27.2876 & -24.6284 & -24.6284 & -14.9258 & -14.9259 &-28.0&-14.6&-24.5275\\
C  & -41.9052   & -41.9053 & -38.0332 & -38.0332 & -23.6568 & -23.6569 &-42.9&-23.0&-37.6863\\
N  & -60.2622   & -60.2623 & -54.9428 & -54.9429 & -34.9084 & -34.9084 &-61.6&-33.9&-54.3977\\
O  & -82.5798   & -82.5799 & -75.5765 & -75.5765 & -48.8831 & -48.8832 &-84.3 &-47.3&-74.8076\\
F  & -109.0592  & -109.0594 & -100.1345 & 100.1346 & -65.7674 & -65.7676 &-111.1&-63.5&-99.4072\\
Ne & -139.8865  & -139.8867 & -128.8014 & -128.8016 & -85.7343 & -85.7344 &-142.3&-82.7&-127.4907\\
\end{tabular}
\end{ruledtabular}
\footnotetext[1] {From Ref.\ \onlinecite{Chan..Handy.2001}. }
\footnotetext[2] {Spin-restricted LDA (Slater exchange) calculation.} 
\end{table*}

\begin{table*}[tb]
\caption{\label{tab:table3}
OF-DFT self-consistent atomic total energies (Hartree a.u.) obtained 
from various kinetic energy functionals and VWN $E_{\rm xc,LDA}$ with 
the numerical grid KS code.
} 
\begin{ruledtabular}
\begin{tabular}{lcccccc}
 & 1/9 vW+TF & 1/5 vW+TF &  vW+TF & GGA (PBE-TW) & mcGGA (PBE2) & KS\footnotemark[1]~ \\
\colrule
H  & -0.7101 & -0.6084 & -0.2924 & -0.76 & -0.36 & -0.4457 \\
He & -3.3244 & -2.9175 & -1.5590 & -3.5  & -1.6 & -2.8348 \\
Li & -8.4175 & -7.4860 & -4.2469 & -8.7  & -4.2 & -7.3352 \\
Be & -16.3982& -14.7162& -8.6995 & -16.9   & -8.5 & -14.4472 \\
B  & -27.5953& -24.9329& -15.2033 & -28.4 &  -14.8 &  -24.3436\\
C  & -42.2886& -38.4132& -24.0078 & -43.3 &  -23.4 &  -37.4202\\
N  & -60.7237& -55.4007& -35.3357 & -62.1 & -34.3 &  -54.0250\\  
O  & -83.1215& -76.1146& -49.3893 & -84.8 & -47.8 &  -74.4613\\ 
F  &-109.6832&-100.7547& -66.3545 & -111.7& -64.1 &  -99.0960\\
Ne &-140.5945&-129.5054& -86.4042 & -143.1 & -83.4&  -128.2335\\
\end{tabular}
\end{ruledtabular}
\footnotetext[1] {Spin-restricted LDA calculation.} 
\end{table*}

These atomic results lead us to nuanced agreement with the
claim of Ref.\ \onlinecite{Chan..Handy.2001} and disagreement with the
claim of Ref.\ \onlinecite{Levy..Sahni84}.  The OF-DFT Euler equation
is not, in general, solvable by simple modification of a standard GTO
KS code (the norm for molecular calculations).  Even a good
all-numerical KS code is challenged to achieve solutions but can be
made to succeed for isolated atoms.  Realizing the computational
speed-up potential of OF-DFT clearly depends on algorithms and
implementations well-suited for OF-DFT, even for one-point functionals.

\subsection{Diatomic Molecules}

Numerical-grid solution of Eq.\ (\ref{II3}) for diatomic molecules, 
if possible, would yield two kinds of insight: 
numerical method behavior and the comparative behavior of $n({\mathbf r})$
and $v_\theta({\mathbf r})$ for different OFKE approximations.  Though
the difficulties of using a modified KS code are just as evident in
this case, we have been able to achieve solutions for several light
molecules.  

Numerical requirements include extremely tight convergence 
tolerances on the eigenvalue $\mu$ and normalization ($10^{-3}$ more 
stringent than normal KS calculations), much larger maximum 
distance cutoff (80 to 100 a.u.\ vs.\ normal KS 30 to 40 a.u.), 
and about a factor of five more points in both of the
prolate spheroidal coordinates (roughly 1100 $\times$ 1300 points vs.\
the typical 200 $\times$ 300).  Even so, the total energy convergence
is mediocre for GGA and mcGGA functionals, about 0.01 Hartree at best. 
Convergence is better for the $T_{\rm TFvW,\alpha}$ functionals, between
0.1 and 1 mHartree. 
This need for extreme measures to achieve  
limited-quality outcomes is an additional confirmation of the unsuitability
of unmodified conventional KS schemes for solutions of the OF-DFT 
Euler equation.

\begin{figure}        
\epsfxsize=7.4cm
\epsffile{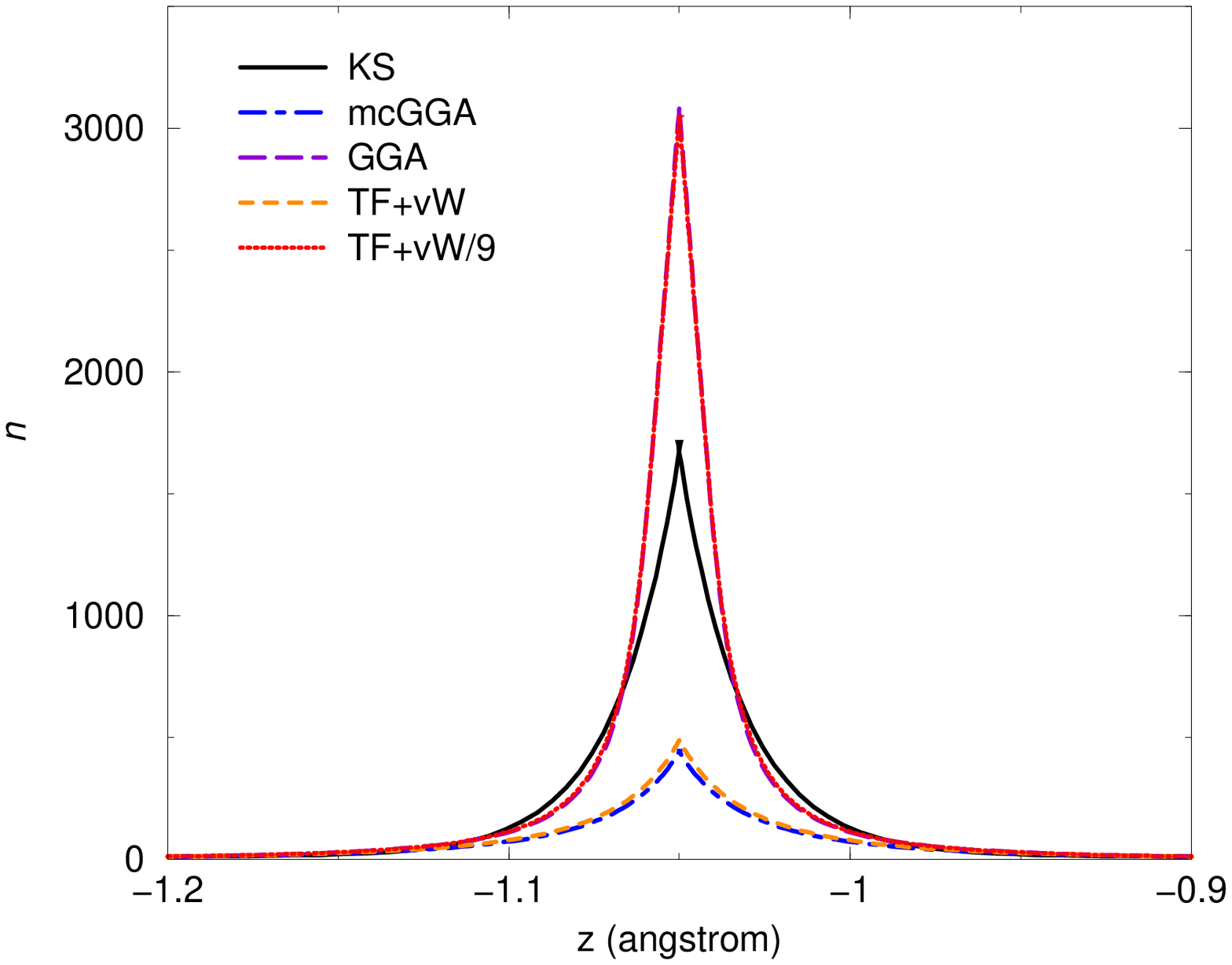}
\caption{
All-electron self-consistent Kohn-Sham and OF-DFT electron densities plotted  
along the SiO molecule internuclear axis in the vicinity of the
Si site.   Si at (0,0,-1.05) \AA, O out of the picture 
at (0,0,+1.05) \AA.  See text.
}
\label{fig2}
\end{figure}

\begin{figure}        
\epsfxsize=7.4cm
\epsffile{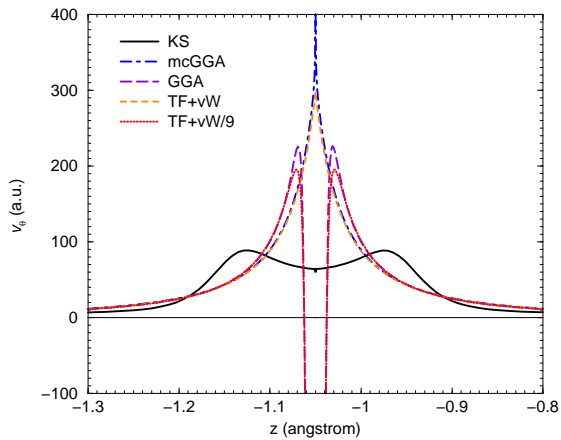}
\caption{
Pauli potentials  $v_\theta$ around the Si site in SiO 
from self-consistent all-electron
Kohn-Sham and OF-DFT calculations with the PBE2 mcGGA, TW 
GGA, and $T_{\rm TFvW,\alpha=1/9,1}$ OFKE functionals. 
}
\label{fig3}
\end{figure}

The solutions nevertheless provide real comparative insight regarding
different OFKE approximations.  Figure \ref{fig2} compares the
all-electron KS density ($E_{\rm xc,LDA}$, VWN) around the Si site in SiO
with the densities from $T_{\rm GGA,PBE-TW}$ (the Tran-Wesolowski
\cite{TranWesolowski02} GGA), $T_{\rm mcGGA,PBE2}$ (the
PBE2 mcGGA \cite{Perspectives}), and
$T_{\rm TFvW,\alpha=1/9,1}$. These approximate functionals form
pairs. $T_{\rm GGA,PBE-TW}$ pairs with $T_{\rm TFvW,\alpha=1/9}$, while
$T_{\rm TFvW,\alpha=1}$ pairs with $T_{\rm mcGGA,PBE2}$.  This pairing conforms to 
the expectations formed in considering the small-$s$ behavior
of the respective enhancement factors. The pairing 
also is interpretable directly from 
the near-nucleus repulsion or attraction behavior of the various 
approximations.  $T_{\rm GGA,PBE-TW}$ generates a $v_\theta$ 
with a spurious
negative singularity near the nuclei,  
while $T_{\rm TFvW,\alpha=1/9}$ drastically lowers the von Weizs\"acker
lower bound to $T_{\rm s}$.  Both lead to excess near-nucleus density.  In
contrast, $v_{\theta,\rm mcGGA,PBE2}$ has spurious positive nuclear site
singularities \cite{Karasiev..Harris.2009}.  Near the nuclei, however,
$v_{\theta,\rm mcGGA,PBE2}$ and $v_{\theta,\rm TFvW,\alpha=1}$ match quite well, 
as shown in Fig.\ \ref{fig3}.  The result, shown in Fig.\ \ref{fig2}, is
that these two functionals give rather close to the same density.  Observe
that the behavior of $v_\theta$  in 
the vicinity of the nucleus for each of these approximate functionals 
differs dramatically from that of $v_\theta$ obtained
by inversion of the standard KS scheme.   It is that improper behavior 
which we believe causes the problems with convergence of standard KS
codes used with approximate OFKE functionals. The positive near-nuclei
singularities appear, in particular, to pose numerical problems.

\begin{figure}        
\epsfxsize=7.4cm
\epsffile{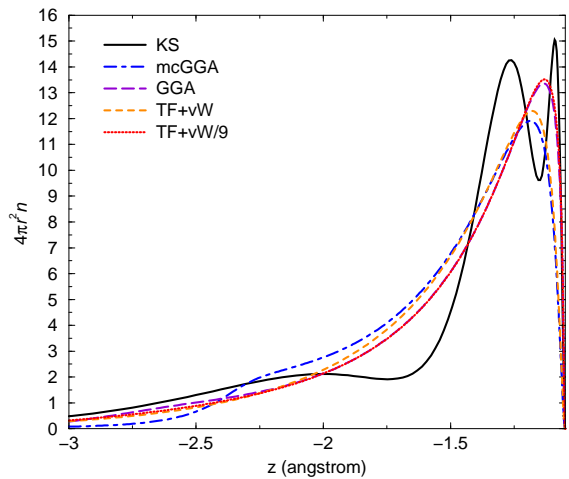}
\caption{
All-electron self-consistent Kohn-Sham and OF-DFT electron densities 
near the Si site along the SiO molecular axis.  These are scaled
scaled by the factor $4\pi(|z|-R/2)^2$,  with $R = 2.10$ \AA, 
the internuclear distance.  This puts the origin of
the scaling at the Si site (0,0,-1.05) \AA. The O is out
of the picture at (0,0,+1.05) \AA.
}
\label{fig4}
\end{figure}

Details of the density near the Si site are provided in Fig.\ \ref{fig4}.
For purposes of display, the densities are weighted by a quasi-radial
factor with origin at the Si site, $4\pi(|z|-R/2)^2$.  The proper
KS shell structure is missing, as is usual with
single-point OFKE functionals.  The more repulsive nature
of the pair $T_{\rm mcGGA,PBE2}$ and $T_{\rm TFvW,\alpha=1}$ compared to 
$T_{\rm GGA,PBE-TW}$ and $T_{\rm TFvW,\alpha=1/9}$ also is evident.  It is 
interesting that in the region $-2.5 < z < 2.0$ $T_{\rm mcGGA,PBE2}$ does 
give a weak mimicry of the outermost shell structure in $T_{\rm s}$, unlike
the other models.  We are uncertain as to how reliable or useful this
feature is.

\begin{figure}        
\epsfxsize=7.4cm
\epsffile{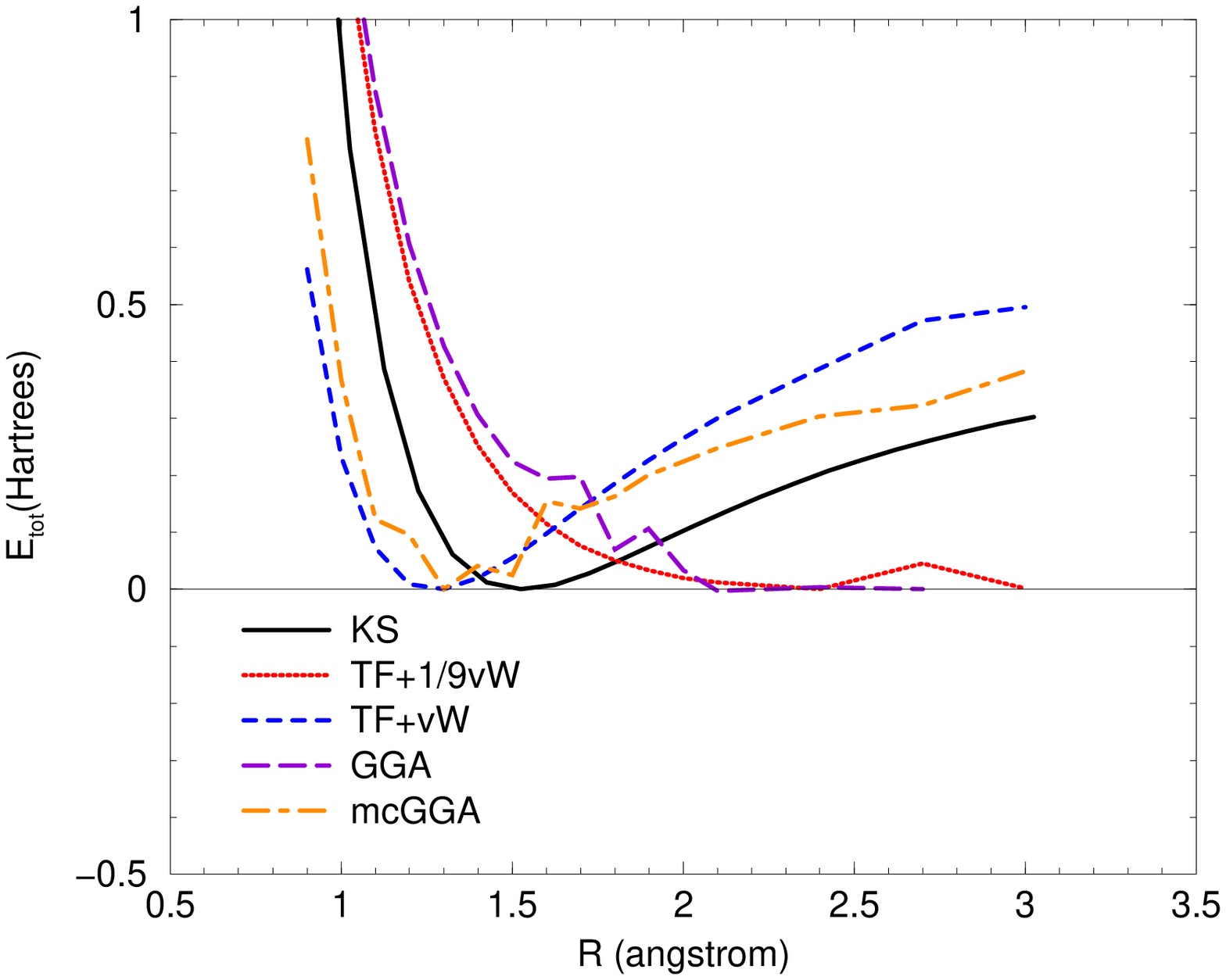}
\caption{
Total energy of the SiO molecule as a function of bond length
obtained from self-consistent all-electron
Kohn-Sham and OF-DFT calculations with Thomas-Fermi, Tran-Wesolowski 
(GGA) and PBE2 (mcGGA) kinetic energy functionals. 
Kohn-Sham values are shown for comparison. 
Values are shifted to a common zero by 363.076 (KS), 386.339 (TF+vW/9), 
250.529 (TF+vW), 83.902 (GGA) and  241.564 (mcGGA) Hartree a.u.
}
\label{fig5}
\end{figure}

Fig.\ \ref{fig5} compares the behavior of $E^{\rm OF-DFT}[n]$ as a 
function of SiO bond length with standard KS results.  
One sees immediately that the
GGA and mcGGA forms introduce numerical difficulties because
of their dependence on the reduced density gradient $s$, Eq. (\ref{sdefn}). 
Clearly there is a grid interval-size problem which could be
obviated by going to even denser grids but at obvious 
computational cost.  The known failure of $T_{\rm GGA,PBE-TW}$ to give binding 
\cite{Perspectives} is evident.  $E_{\rm TFvW,\alpha=1/9}$ 
apparently does not bind either, in keeping with the too-weakened 
lower bound just discussed.  
Full TFvWD and $E_{\rm mcGGA,PBE2}$ are fairly close, with the mcGGA being
the best of the lot with respect to equilibrium bond length.   

\subsection{Simple Analysis of the Difficulty}

The barrier to use of a standard KS code to solve Eq.\ (\ref{II3}) 
can be traced to the near-nucleus repulsion of $v_\theta$.  As displayed
in Fig.\ \ref{fig3}, the exact 
$v_\theta$ is strongly repulsive in a fairly small region around the
nuclear site.  Ref.\ \onlinecite{Karasiev..Harris.2009}, Fig.\ 2 shows
that the exact $v_\theta$ can  have rather sharp structure within a 
radius of about 1 bohr of a nuclear site.  In contrast, some simple 
approximations which are properly positive definite, including our mcGGA, 
actually are singular at the nuclei; again see Fig.\ \ref{fig3}.  Such 
strong repulsion overwhelms the attractive $v_{\mathrm{xc}}$.  That figure
also shows that some approximations deliver Pauli potentials with 
negative nuclear-site singularities.  We consider 
that case below.  First, however, the simplest example
will suffice to illustrate the problem with properly positive 
$v_\theta$.  Pick $T_\theta = T_{\mathrm{TF}}$, Eq.\ (\ref{I6}), 
and $E_{\mathrm{xc}}$ to be simplest Slater exchange:
\begin{align}
E_{\mathrm {xc}} =& c_{\mathrm x}  \int n^{4/3} ({\mathbf r}) %
d^3{\mathbf r} \,  \nonumber \\
c_{\mathrm x} =& -\frac{3}{4}\left(\frac{3}{\pi}\right)^{1/3} 
\label{ExcSlater}
\end{align}
Then in Eq.\ (\ref{II3}), the potentials become 
\begin{align}
v_\theta =& \half (3 \pi^2)^{2/3}n^{2/3}({\mathbf r})\,  \nonumber \\
v_{\mathrm{KS}} =& v_{\mathrm H} + v_{\mathrm{Ne}} + v_{\mathrm{xc}} \nonumber \\
v_{\mathrm{xc}} =& -\left(\frac{3}{\pi}\right)^{1/3}n^{1/3}({\mathbf r})\,  \nonumber \\
\label{potentials}
\end{align}
A hydrogen-like density,
\be
n_{\mathrm H}({\mathbf r}) := \frac{N_e^4}{\pi}\exp(-2N_e r) \; ,
\label{Hdens}
\ee
obeys the Kato cusp condition near the nucleus 
\cite{Kato57,Bingel63,PackByersBrown66,March..VanDoren2000,KryachkoLudena}, 
hence is useful for testing.  At the nucleus, this density yields 
the ratio of potentials 
\be
\frac{v_{\mathrm{xc}}(0) + v_\theta(0)}{|v_{\mathrm{xc}}(0)|} = %
-1 + 3.318004 N_e^{4/3}  \;. 
\label{potratio}
\ee
For $N_e = 6$, this ratio is already 35.2.  By 
$N_e =10$ it is 70.5. Additional simple calculations with 
the potential which appears in Eq. (\ref{II3}) without the positive 
Hartree contribution, that is $v_{\rm Ne}(r)+v_\theta(r)+v_{\mathrm{xc}}(r)$,
illustrate the point.  
At small $r$ with $N_e=6$, that potential becomes positive for $r>0.028$ Bohr. 
For $N_e=10$ it is positive for $r>0.011$ Bohr. 
These little exercises illustrate why the use of an ordinary K-S code
becomes so difficult.  Such peculiar  
behavior is quite different from what is encountered in the ordinary 
KS problem.  

{F}rom the perspective of numerical stability,
the case of a $v_\theta$ which has a spurious negative singularity 
near each nuclear site, {\it e.g.} PBE-TW, is at least as bad if not
worse.  As a site is approached, such potentials first are increasingly
repulsive, then  plunge abruptly into the negative singularity; see
Fig.\ \ref{fig3}.  

\section{Pseudo-potential Solutions of the OF-DFT Euler Equation}

Having demonstrated the difficulties with solving the OF-DFT Euler-Lagrange
problem with a modified KS eigenvalue code, we turn to the use of
direct Euler-Lagrange minimization. 
The specific objective is to exploit the numerical methodology in the
{\sc Profess} code \cite{Ho..Carter08,Hung..Carter10}.  Written
originally for use with two-point functionals, {\sc Profess} performs
OF-DFT calculations by minimization
of the Euler-Lagrange equation as a functional of $n({\mathbf
  r})$ under periodic boundary conditions.  It uses a numerical 3D
mesh and FFTs.  
As published, the code includes the TF, vW, and ${\rm TFvW,\alpha}$ functionals
as well as the Wang-Teter (WT) \cite{Wang.Teter.1992}, and
Wang-Govind-Carter (WGC) \cite{Wang.Govind.Carter.1998} OFKE functionals.  The
PZ and PBE $E_{\rm xc}$ functionals are implemented in 
{\sc Profess}.  For this study, we added the Tran-Wesolowski GGA  
\cite{TranWesolowski02} and our PBE2 mcGGA \cite{Perspectives} 
OFKE functionals.  

As is the case with standard KS calculations done in a plane-wave
basis, {\sc Profess} relies upon 
pseudo-potential (PP) techniques to screen the nuclear-electron
potential cusp and exclude chemically inactive core states. 
Though OF-DFT has no problem with core states and the density (and 
its square root) is a comparatively 
unstructured, smooth function, regularization of the nuclear-electron
interaction singularity still is a requisite for an efficient implementation.  

High-quality pseudo-potentials developed for conventional KS
calculations generally are non-local, in the specific sense that they
contain projection operators which provide different potentials for
different orbital angular momenta.  That explicit orbital dependence
makes non-local pseudo-potentials (NLPP) inapplicable in OF-DFT
calculations. Instead, local pseudo-potentials (LPP), {\it i.e.}, of
the form of a simple multiplicative operator which is the same for all
orbitals, must be developed.  {\sc Profess} requires a LPP in real or
reciprocal space as input. Observe that this limitation to local form
is an additional approximation, over and beyond the PP itself, which
has accuracy limitation implications for both conventional,
orbital-based KS or OF-DFT implementations.

In addition to their simplicity, there is a  
formal advantage of LPPs which is at least of peripheral interest
here. Calculations with local PPs are within the framework of the 
standard KS scheme, which assumes a local effective potential. 
The NLPP case obviously does not meet that assumption.  Although the 
Hohenberg-Kohn theorem has been extended to the case of a non-local external 
potential \cite{Gilbert.1975}, the exchange-correlation energy in that case
becomes be a functional of the one-particle reduced density matrix 
instead of a functional $n({\mathbf r})$ alone.  

Many methods have been proposed to develop LPPs. Among them we mention
(i) empirical (or model) LPPs as, for example in
Refs. \onlinecite{Shaw.1968,Topp.Hopfield.1973,Heine.Abarenkov.1964,Goodwin..Heine.1990, %
  Fiolhais..Brajczewska.1995}; (ii) local potentials obtained from
non-local ones, for example, by use of just one $l$-channel from an NLPP 
as a LPP (for example, Ref.\ \onlinecite{Cabral.Martins.1995}; 
(iii) LPPs constrained to reproduce
atomic properties, eigenvalues, or pseudo-density, etc., which follow
from  a (presumably superior) NLPP (for example, Ref.\ \onlinecite{ALPS-BLPS}), 
and finally (iv) local PPs 
derived to reproduce some bulk property values, either experimental or
those  predicted by NLPP calculations 
\cite{Watson..Carter.Madden.1998,Huang.Carter.2008,ALPS-BLPS}

\subsection{Local Pseudo-potentials for OF-DFT Calculations}

\subsubsection{Development}

Some time ago, an iterative procedure was developed 
\cite{Baerends-DD.1994,Baerends-DD.1995}  to 
solve the inverse problem of determining the KS 
effective potential $v_{\rm KS}({\mathbf r})$  
from a given density $n({\mathbf r})$.  
Subsequently, we \cite{Karasiev..Harris.LPP}  introduced
and tested an improvement.  In the case of Li, however, 
both versions share a problem. For a single valence orbital 
(singly or doubly occupied) the solution of the inverse problem 
is trivial and known.  The local pseudo-potential is equal 
to the $s$-channel of the NLPP, $v_{local}(r)=v_{l=0}(r)$.  
Hence the LPP contains no information about the $l>0$ channels 
of the NLPP.  Those channels are critical in crystalline binding.

Therefore,  to include information about all $l$ channels of
the reference NLPP, we consider a sort of normalized linear combination 
of $l$ components of that NLPP, 
\begin{equation}
v_{l_{max}}({\mathbf r})=\sum_{l=0}^{l_{max}} c_l v_l({\mathbf r}) %
\Big/\sum_{l=0}^{l_{max}} c_l \;
\label{E2}
\end{equation} 
where the parameters $\{c_l\}$ are to be adjusted to fit selected
equilibrium bulk material properties calculated with the reference KS method.   
This particular method of LPP generation amounts 
to a mixture of methods (ii) and (iv) described at the outset of 
this Section.

In the present case, we simply took the 
bcc Li lattice constant as predicted by a standard KS
calculation with PBE \cite{PBE} $E_{\rm xc}$, and the
plane wave (PW) basis set (see Table \ref{tab:table4}), namely  $a=3.44$~\AA.
Components of the Troullier-Martins norm-conserving NLPP were 
used in Eq. (\ref{E2}).
For generation of the NLPP with PBE XC, 
we took the core radius to be 2.45 a.u. 
The parameters $\{c_l\}$ in Eq.\ (\ref{E2}),
for the $s$, $p$, and $d$ channels
respectively, were determined by constraining a KS calculation with 
the LPP Eq.\ (\ref{E2}) to 
reproduce the reference optimized bcc Li lattice constant value. 
Those KS  calculations  done 
with  PBE XC 
in the {\sc Siesta} code \cite{siesta} and a DZP numerical
atomic orbital (NAO) basis set. 
The optimized parameter values are
 $c_0=0.69$, $c_1=0.34$, $c_2=0.10$.  We designate this LPP 
as $v_{\rm GGA, spd1}$.  
To generate the LDA local LPP, $v_{\rm LDA,spd1}$
for the Perdew-Zunger \cite{PZ81} LDA XC functional, 
components of the LDA NLPP and the 
same set of the channel-mixing parameters were used in Eq. (\ref{E2}).

An alternative LPP form which we also studied is a modification of 
the potential proposed by Heine and Abarenkov 
\cite{Heine.Abarenkov.1964,Goodwin..Heine.1990}. In real space, 
the Heine-Abarenkov 
model potential is
\begin{equation}
v_{\rm mod}(r) = \left\{ \begin{array}{ll}
-A,   & r< r_c \\
-Z/r, & r\ge r_c\\
\end{array} \right.
\label{E3}
\end{equation}
where $A$ is a constant, $r_c$ is the core radius, and $Z$ is the core charge. 
The model potential in reciprocal space is given by
\begin{equation}
v_{\rm mod}(q)=\frac{-4\pi}{\Omega q^2}[(Z-Ar_c){\rm cos}(qr_c)+(A/q){\rm sin}(qr_c)],
\label{E4}
\end{equation}
where $\Omega$ is the unit cell volume. In
Ref.\ \onlinecite{Goodwin..Heine.1990}, this potential was 
multiplied by a smoothed step
function $f(q)={\rm exp}[-q/q_c)^6]$ to reduce spurious oscillations
in $v_{\rm mod}(q)$ and to ensure rapid decay of $v_{\rm mod}(q)$ at large
wave-vectors.  Those oscillations are caused by the discontinuity 
of the real-space
potential at the core radius.  Here, the parameter $q_c$ was chosen 
as suggested in Ref.\ \onlinecite{Goodwin..Heine.1990}, namely, 
to equal the second zero position of $v_{\rm mod}(q)$.

To obtain counterparts 
of the local potentials $v_{\rm GGA,spd1}$ and $v_{\rm LDA,spd1}$,
Eq.\ (\ref{E3})
in the simple modified Heine-Abarenkov model form,
the two parameters, $A$ and $r_c$, were determined by minimization of
$\int d{\mathbf r}|v_{\rm spd1}({\mathbf r})-v_{\rm mod1}({\mathbf r})|^2 $. 
This yields $A=0.45499$ Hartrees, $r_c=2.2261$ Bohr, $g_c=2.86$ Bohr$^{-1}$ for
$v_{\rm GGA,mod1}$ and $A=0.45376$ Hartrees, $r_c=1.8818$ Bohr, $g_c=2.94$ Bohr$^{-1}$ for
$v_{\rm LDA,mod1}$. 
The local potentials $v_{\rm GGA,spd1}$ and $v_{\rm LDA,spd1}$ in 
reciprocal space are
multiplied by the same smoothed step function $f(q)$ with $q_c$ values equal to
2.95 and 3.25 Bohr$^{-1}$ respectively.

\begin{figure}        
\epsfxsize=7.4cm
\epsffile{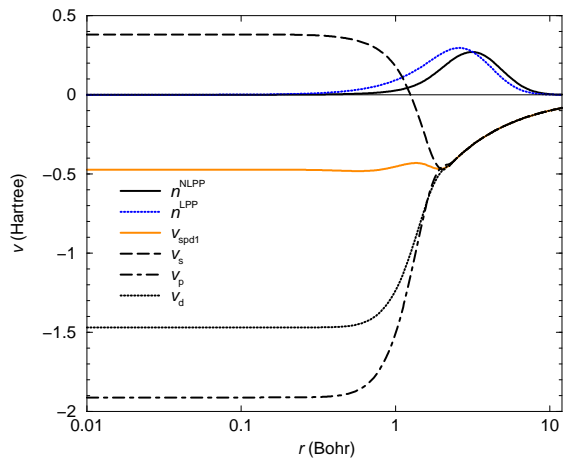}
\caption{
Real space pseudo-potentials for Li: local $v_{\rm GGA,spd1}$, and 
different $l$-components of the non-local Troullier-Martin (TM) 
pseudo-potential. 
Pseudo-densities generated with local and non-local PPs are shown for 
comparison. 
}
\label{fig_pp1}
\end{figure}

\begin{figure}        
\epsfxsize=7.4cm
\epsffile{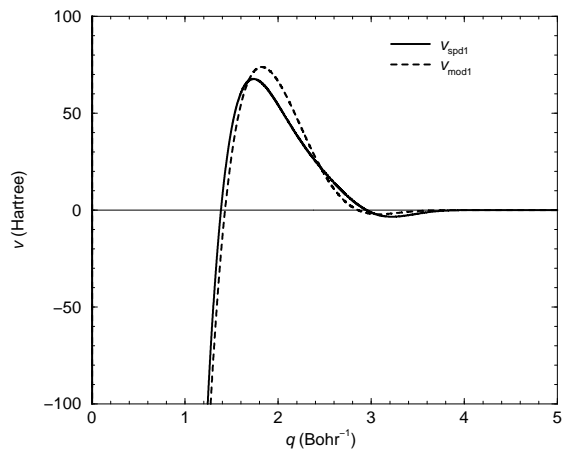}
\caption{
Reciprocal space local pseudo-potentials for Li: 
$v_{\rm GGA,spd1}$ 
and 
$v_{\rm GGA,mod1}$. 
}
\label{fig_pp2}
\end{figure}

Figure \ref{fig_pp1} shows the $v_{\rm GGA,spd1}$ LPP in real space in
comparison with the NLPP $l$ channels, along with the two
pseudo-densities which result.  Figure \ref{fig_pp2} shows the $v_{\rm
  GGA,spd1}$ and $v_{\rm GGA,mod1}$ LPPs in reciprocal space.

\subsubsection{KS Tests of Local Pseudo-potentials}


Kohn-Sham calculations with the LPPs  
were performed using the {\sc Abinit} PW code \cite{AbInit} 
with PZ and PBE exchange-correlation
functionals.  We also used {\sc Siesta} with the same exchange-correlation
functionals and a 2s$^2$2p$^2$ numerical atomic orbital basis set 
(8 NAO per atom).
Table \ref{tab:table4} shows the equilibrium lattice constants and bulk moduli
for the various LPPs. Those results are compared to the Kohn-Sham calculations 
performed with the non-local projector augmented wave (PAW) scheme 
(as implemented in {\sc Vasp} and {\sc Abinit}) and 
TM norm-conserving pseudo-potentials with core correction \cite{TM-abinit}. 
The lattice constant and bulk modulus reported in 
Table \ref{tab:table4} were obtained
by fitting the calculated total energies per cell to 
the stabilized jellium model equation of state (SJEOS, \cite{SJEOS.2001}).
All the local PPs reproduce 
the PAW results rather closely for both lattice constant 
and bulk modulus. The bulk moduli 
calculated using NAO orbitals and norm conserving TM pseudo-potentials 
are slightly larger than the PAW plane wave results.

As a check against an all-electron localized-orbital calculation,
we did high-quality GTO-basis KS calculations (10s6p3d basis) with the GTOFF 
code\cite{TAB}.  For $E_{\rm xc,PZ}$ and $E_{\rm xc,PBE}$, we obtained 
optimized bcc Li lattice parameters of 3.360 and 3.435 \AA, respectively, 
essentially the same as from the {\sc Siesta} NAO and plane wave PAW calculations.  

%
\begin{table}
\caption{\label{tab:table4}
Kohn-Sham lattice constant ($\AA$) and bulk modulus (GPa) for bcc Li 
calculated 
using {\sc Vasp} PW PAW schemes, {\sc Abinit} PW PAW and local 
pseudo-potentials, 
{\sc Siesta} non-local Troullier-Martins \cite{TM-abinit} 
and local pseudo-potentials.
Orbital-free calculations used  
$T_{\rm mcGGA,PBE2}$, $T_{\rm TFvW,\alpha=1}$,
$T_{\rm GGA,PBE-TW}$, and $T_{\rm TFvW,\alpha=1/9}$ kinetic energy functionals 
in combination with $E_{\rm xc,LDA,PZ}$ and $E_{\rm xc,GGA,PBE}$ with local 
pseudo-potentials $v_{\rm LDA,spd1}$, $v_{\rm GGA,spd1}$, $v_{\rm LDA,mod1}$ 
and $v_{\rm GGA,mod1}$. Conventional KS calculations were done with 
a 2-atom unit cell
and 7 $\times$  7 $\times$  7 ({\sc Vasp} and {\sc Siesta})
or 9 $\times$  9 $\times$  9 ({\sc Abinit}) $\mathbf k$-mesh.  The {\sc Siesta} 
basis set was 2s$^2$2p$^2$ (8 NAO per atom).
Orbital-free calculations used a  
128-atom supercell.}
\begin{ruledtabular}
\begin{tabular}{lccccc}
&&\multicolumn{2}{c} {LDA} & \multicolumn{2}{c} {$\rm GGA$}\\
\cline{3-4}\cline {5-6}
Method & PP &  $a$  & $B$  & $a$  & $B$ \\
\hline
\underline{Kohn-Sham} &&&&&\\
PW ({\sc Vasp})  &PAW    & 3.37 & 15.0 & 3.45 & 13.7 \\
PW ({\sc Abinit})&PAW    & 3.37 & 15.1 & 3.44 & 13.9 \\
NAO ({\sc Siesta})&TM     & 3.37 & 15.6 & 3.44 & 14.3 \\
\hline
\underline{Kohn-Sham} &&&&&\\
PW ({\sc Abinit}) &spd1\tablenotemark[1]   & 3.37 & 14.8 & 3.44 & 13.8 \\
NAO ({\sc Siesta})&spd1\tablenotemark[1]   & 3.38 & 14.9 & 3.45 & 13.9 \\
\hline
\underline{Kohn-Sham} &&&&&\\
PW ({\sc Abinit})&mod1\tablenotemark[2]   & 3.37 & 14.8 & 3.44 & 13.9 \\
NAO ({\sc Siesta})&mod1\tablenotemark[2]   & 3.38 & 14.9 & 3.44 & 13.9 \\
\hline
\hline
\underline{OFDFT} &&&&&\\
mcGGA &spd1\tablenotemark[3]   & 3.37 & 16.2 & 3.43 & 15.4 \\
TF+vW &spd1\tablenotemark[3]   & 3.37 & 16.0 & 3.43 & 15.2 \\
GGA   &spd1\tablenotemark[3]   & 3.37 & 11.8 & 3.46 & 11.8 \\
TF+1/9vW &spd1\tablenotemark[3]& 3.37 & 11.4 & 3.46 & 11.4 \\
\hline
\underline{OFDFT} &&&&&\\
mcGGA  &mod1\tablenotemark[4]   & 3.36 & 16.2 & 3.43 & 15.2 \\
TF+vW &mod1\tablenotemark[4]    & 3.37 & 15.9 & 3.43 & 14.9 \\
GGA   &mod1\tablenotemark[4]    & 3.42 & 10.8 & 3.49 & 10.1 \\
TF+1/9vW &mod1\tablenotemark[4] & 3.42 & 10.3 & 3.49 & 9.5 \\
\end{tabular}
\tablenotetext[1]{Real space potential defined by Eq. (\ref{E2}) (see text for details).}
\tablenotetext[2]{Real space potential defined by Eq. (\ref{E3}) (see text for details).}
\tablenotetext[3]{Reciprocal space potential defined by Fourier-Bessel transform of local
potential Eq. (\ref{E2}) and multiplied by $f(q)$ function (see text for details).}
\tablenotetext[4]{Reciprocal space potential defined by Eq. (\ref{E4}) multiplied by $f(q)$ function (see text for details).}
\end{ruledtabular}
\end{table}

\subsection{Pseudo-potential OF-DFT Tests}

\subsubsection{OF-DFT Comparison for bcc Li}

\begin{figure}        
\epsfxsize=7.4cm
\epsffile{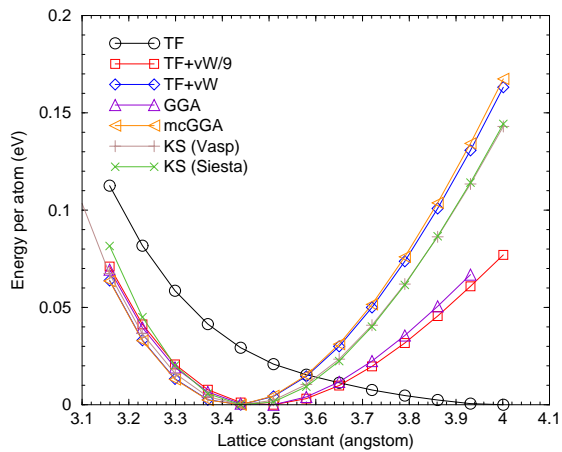}
\caption{
Energy per atom vs.~lattice constant for bulk bcc Li. OF-DFT results 
for $T_{\rm TF}$, $T_{\rm TFvW,\alpha}$, $T_{\rm GGA,PBE-TW}$, and 
$T_{\rm mcGGA,PBE2}$ compared 
to the KS values.  OF-DFT calculations with 
128-atom supercell, $v_{\rm GGA, mod1}$ LPP, and $E_{\rm xc,GGA,PBE}$.  
KS calculations 2-atom unit cell with non-local PAW PBE pseudo-potentials ({\sc Vasp}) 
and with Troullier-Martin PPs with PBE exchange-correlation, 
$\rm 2s^22p^2$ basis set (8 NAO per atom) ({\sc Siesta}). 
}
\label{fig_pp3}
\end{figure}

For the OF-DFT bcc Li studies, we used a 128-atom supercell 
in {\sc Profess} with 
the $v_{\rm spd1}$, $v_{\rm mod1}$  LPPs just described and  both 
$E_{\rm xc,LDA}$ and $E_{\rm xc,GGA}$.  
We did the {\sc Profess} calculations for the 
 $T_{\rm TF}$, $T_{\rm TFvW,\alpha=1, 1/9}$, $T_{\rm GGA,PBE-TW}$, and 
$T_{\rm mcGGA,PBE2}$ functionals.  
The computed $E_{\rm tot}/$atom values are plotted as a function
of bcc lattice constant in Fig. \ref{fig_pp3}.

One sees that, as might be expected, the pure TF+XC model fails to bind. The
pairing of other functionals, which we have discussed already, 
reappears. 
$T_{\rm TFvW,\alpha=1}$ pairs with $T_{\rm mcGGA,PBE2}$, and  
$T_{\rm TFvW,\alpha=1/9}$ pairs with $T_{\rm GGA,PBE-TW}$.  The former pair 
gives a better 
description of both the lattice constant and bulk modulus than the 
latter pair.  The computed equilibrium lattice constants and bulk moduli
are shown in Table \ref{tab:table4}.

The equilibrium lattice constants 
predicted by the OF-DFT calculations with $v_{\rm spd1}$ LPPs 
agree well with the KS PAW results. When the $v_{\rm mod1}$ model
pseudo-potential is used, the lattice constant  from 
the OF-DFT calculations with $T_{\rm GGA,PBE-TW}$ and 
$T_{\rm TFvW,\alpha=1, 1/9}$ is an over-estimate of about 1 \% 
for both LDA and GGA XC functionals.  
This pair of OFKE functionals also predicts low bulk modulus
values, again for both LDA and GGA XC cases. The mcGGA and TF+vW 
KE functionals do very well for the lattice parameter and 
slightly overestimate the bulk modulus value.

\begin{figure}        
\epsfxsize=7.4cm
\epsffile{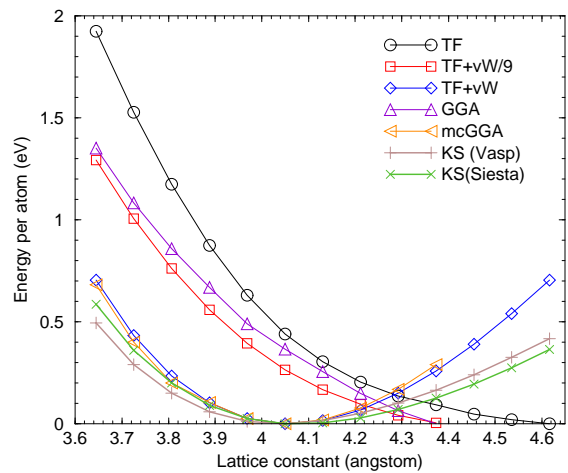}
\caption{
As in Fig.\ \ref{fig_pp3} for bulk Al but with a 
4-atom supercell. {\sc Siesta} calculations performed with standard DZP basis set.
The Goodwin, Needs, and Heine 
\cite{Goodwin..Heine.1990}  local model pseudo-potential used in the orbital-free calculations.  See text.
}
\label{fig_pp4}
\end{figure}
  
\subsubsection{OF-DFT Comparison for fcc Al}

The utility of existing LPPs for OFDFT calculations obviously is 
a pertinent issue.  To explore that, we considered bulk Al.  
The model LPP in the form of Eq.\ (\ref{E4}) with
parameters from Goodwin, Needs, and Heine
\cite{Goodwin..Heine.1990} was used in OF-DFT calculations.  As
before, this was done with the  
five OF-KE functionals, but here only in combination with 
the PBE GGA XC functional.
Fig.\ \ref{fig_pp4} shows {\sc Profess} results for a 4-atom fcc cell 
compared with conventional KS results obtained with 
{\sc  Vasp} in the same cell with a 5 $\times$  5 $\times$  5 
$\mathbf k$-mesh calculation.
The GGA and mcGGA KE functionals introduce numerical instability
at expanded geometry. Aside from that, one again 
observes the same pairing of KE functionals as before.  
The $T_{\rm GGA,PBE-TW}$ and $T_{\rm TFvW,\alpha=1/9}$ functionals
do not produce detectable minima. 
The $T_{\rm mcGGA,PBE2}$ and $T_{\rm TFvW,\alpha=1}$
pair predict equilibrium lattice constants
($a=4.05$ and 4.06 $\AA$ correspondingly), very close to the KS results
($a=4.05$ and 4.09 $\AA$ for PAW {\sc Vasp} and NAO DZP {\sc Siesta} 
calculations respectively). However, 
the shape of the two OF-DFT energy curves differs perceptibly from
the KS results.  In particular, the OF-DFT functionals predict a
softer solid.  

\section{Summary Discussion}

Several clear results emerge from this study.  First, use of standard
KS codes to solve the OF-DFT Euler equation as a modified KS
eigenvalue problem is problematic at best.  At least for the
all-electron case, it seems implausible as a productive route to
routine OF-DFT calculations.  One could speculate that
a better-behaved one-point approximate OFKE functional than mcGGA might not
be such a challenge to standard KS algorithms.  The repulsive nature
of even the exact $v_\theta$ (recall Fig. \ref{fig3}) makes that outcome
seem rather doubtful.  

Second, even if a particular approximate one-point OFKE functional has
singular behavior, it is possible that such a functional can deliver
physically realistic results.  Those results can be obtained with a
sufficiently refined direct Euler-Lagrange solution of the effective
KS equation, Eq.\ (\ref{A3}).  Thus, we are able to extract useful,
self-consistent solutions for the recently developed simple mcGGA
OFKE functional as well as the Tran-Wesolowski GGA.  These
solutions enable understanding of the consequence 
of the singular behavior of their respective Pauli potentials. The
Tran-Wesolowsk GGA has attractive singularities which cause strong
over-estimates of the self-consistent density near the nuclear
sites. In contrast, the properly positive mcGGA OFKE Pauli potential
has positive singularities near the nuclei and the density is 
underestimated there.

Third, we have presented a procedure for developing a local pseudo-potential
for OFDFT calculations by doing a multi-channel weighting of a
corresponding non-local pseudo-potential.  The weighting is
determined by KS calculations with the LPP such that the equilibrium
non-LPP lattice parameter is reproduced.  We showed that this yields
a very good LPP.   A remaining challenge for the
OFDFT agenda is to construct a good LPP from an existing non-LPP without
appeal to any bulk or aggregate system KS calculations.

Fourth, once a suitable local pseudo-potential procedure is 
defined, the progress made on computational solution of the minimization problem
for two-point OFKE approximations can be appropriated directly for 
use with one-point OFKE approximations.  Even so, we do observe
numerical instabilities in the case of the  mcGGA and GGA
OFKE functionals.

\begin{acknowledgments}
We acknowledge informative conversations with Frank Harris, 
Travis Sjostrom, and Jim Dufty with thanks. This work 
was supported in part by the U.S.\ Dept.\ of Energy TMS program,  
grant DE-SC0002139.
\end{acknowledgments}


\begin{thebibliography}{99}
\bibitem{TM-abinit}
Pseudo-potentials are taken from the {\sc Siesta} repository
of translated norm-conserving Troullier-Martins pseudo-potentials from
the Fritz-Haber-Institute.
\end{thebibliography}
\end{document}